\documentclass[twocolumn,pre,showpacs]{revtex4}

\usepackage{graphicx,amssymb,amsmath}

\begin{document}

\title{Nonlinear dynamics of flexural wave turbulence}
\author{Benjamin Miquel}
\affiliation{Laboratoire de Physique Statistique, Ecole Normale Sup\'erieure, Universit\'e Pierre et Marie Curie, CNRS, 24 rue Lhomond, 75005 Paris, France.}
\author{Nicolas Mordant}
\email[]{nicolas.mordant@legi.grenoble-inp.fr}
\affiliation{Laboratoire des Ecoulements G\'eophysiques et Industriels, CNRS/UJF/G-INP, BP53, 38041 Grenoble, France}
\affiliation{Institut Universitaire de France}


\begin{abstract}
The Kolmogorov-Zakharov spectrum predicted by the Weak Turbulence Theory remains elusive for wave turbulence of flexural waves at the surface of an thin elastic plate. We report a direct measurement of the nonlinear timescale $T_{NL}$ related to energy transfer between waves. This time scale is extracted from the space-time measurement of the deformation of the plate by studying the temporal dynamics of wavelet coefficients of the turbulent field. The central hypothesis of the theory is the time scale separation between dissipative time scale, nonlinear time scale and the period of the wave ($T_d>>T_{NL}>>T$). We observe that this scale separation is valid in our system. The discrete modes due to the finite size effects are responsible for the disagreement between observations and theory. A crossover from continuous weak turbulence and discrete turbulence is observed when the nonlinear time scale is of the same order of magnitude as the frequency separation of the discrete modes. The Kolmogorov-Zakharov energy cascade is then strongly altered and is frozen before reaching the dissipative regime expected in the theory.
\end{abstract}

\maketitle

\section{Introduction}

The nonlinear coupling among a large number of waves causes an energy transfer between them and yields a state of wave turbulence. This phenomenon plays an important role in the atmosphere/ocean coupling through surface waves forced by the wind, in energy transport and dissipation by internal waves across the ocean, energy exchanges by Rossby waves in the atmosphere, energy dissipation by Kelvin waves in superfluid turbulence or confinement in tokamak plasmas among many examples. The weak turbulence theory (WTT) aims at developing a theoretical framework for such turbulence of dispersive waves~\cite{Zakharov, Newell,Nazarenko}. It relies on two major assumptions: weak nonlinearity and an asymptotically large size of the system. Because of the weak nonlinearity, energy exchanges between waves are slow and restricted to resonant waves that conserve energy and impulsion. The wave amplitude is slowly modulated by the coupling and an energy cascade operates from the large scales of the energy input to the small scales at which it is dissipated. The theory relies fundamentally on the following scale separation 
\begin{equation}
T\ll T_{NL}\ll T_d
\end{equation}
where $T=2\pi/\omega$ is the period of the wave, $T_{NL}$ the characteristic time of the nonlinear exchange of energy in the cascade and $T_d$ is the characteristic time of dissipation. Both latter times usually depend on $\omega$. Thanks to this scale separation, an asymptotic calculation can be performed to predict the statistical properties of wave turbulence~\cite{Zakharov,Newell, Nazarenko}. It predicts in particular an energy cascade in scales sharing many similarities with the Kolmogorov cascade of hydrodynamical turbulence (thus the name Kolmogorov-Zakharov cascade for wave turbulence). The goal of this article is to check directly the hypothesis of scale separation in flexion wave turbulence in an elastic plate.

Testing the validity of WTT in experiments is challenging because a measurement resolved both in space and time is required in order to probe the structure of the turbulent wave field. In this respect, the time resolved profilometry technique applied to water surface waves and waves on a thin elastic plate is a very valuable tool~\cite{Cobelli,Cobelli1}. In the case of water waves it was observed that the waves develop strongly nonlinear structures (crests and bound waves) that explain the disagreement of the observed wave spectra with the WTT predictions~\cite{Herbert,falcon,denissenko}: for instance the spectral exponents of the wave spectrum predicted by the WTT are not observed in experiments. The case of waves in a thin elastic plate is somewhat different: the space-time spectra of such turbulence show that the waves are indeed weakly nonlinear~\cite{Cobelli, epjb, Miquel}: the dispersion relation remains as a single energy line in the $(k,\omega)$ space (as opposed to water waves) and it is weakly shifted from the linear dispersion relation. Despite this agreement with the WTT requirement of weak nonlinearity, the observed wave spectra are not in agreement with the theoretical predictions~\cite{During,Boudaoud,Mordant}. This disagreement between experiment and theory may be attributed to two effects. First, the forcing is localized in space and thus requires some propagation distance to randomize the wave phases and to develop the weak turbulence energy cascade~\cite{Miquel}. This effect impacts the observed spectrum because the measurement region cannot be located far enough from the forcing in a region where the stationary energy cascade is effective. A second effect may be in action due to the finite size of the plate: for finite systems, the modes have discrete frequencies. This discreteness may hinder the occurrence of resonances if the separation between frequencies is so large to prevent resonance conditions to be fulfilled~\cite{Kartashova}. Assuming a system of size $L$, the wavenumbers of the neighboring modes should be separated by $\delta k\propto \pi/L$. This translates through the group velocity $v_g$ in a frequency separation $\delta \omega_{sep}= v_g(\omega) \delta k$. Such modes have a spectral width $\Delta \omega$ due to dissipation (then $\Delta \omega\propto 1/T_d$ which is a lower bound of the spectral width) and to the nonlinear exchanges of energy that alter the temporal coherence of the waves. In such a case one expects an additional contribution proportional to $1/T_{NL}$ which dominates over the dissipation if the scale separation $T_{NL}\ll T_d$ is fulfilled. The WTT assumes that the frequencies can vary continuously: this is valid if the spectral width of the modes is larger than their separation ($\Delta \omega\gg\delta \omega_{sep}$). The discreteness of the modes can be neglected if $1/T_{NL}\gg\delta \omega_{sep}$. If this is not the case, the separation of the modes prevents the Kolmogorov-Zakharov (KZ) cascade to develop. The dynamics of such discrete wave turbulence resembles rather that of a finite dimensional nonlinear system~\cite{Kartashova}. In such a system, the energy transfer to small scales is strongly affected compared to the KZ cascade. The continuous KZ cascade is expected to be observed in the range of frequencies where
\begin{equation}
T\ll T_{NL}\ll 1/\delta \omega_{sep}\, .
\end{equation}
Depending of the ordering of $T_d$ and $1/\delta \omega_{sep}$ a range of discrete turbulence could be observed as well.

The experimental setup is similar to the one described in~\cite{epjb}: a stainless plate ($1m\times 2m\times 0.4mm$) hangs vertically, clamped at the top on its short side. An electromagnetic shaker is attached slightly above the inferior edge of the plate (see fig.~\ref{exw}(b)) driven at $30\ \mathrm{Hz}$ with various amplitudes. A videoprojector displays a line pattern on the plate. The deformation of the plate yields a deformation of the pattern. This deformed pattern is recorded by an high speed camera at a frame rate up to $10,000$ frames/s. The deformation field $\zeta(\mathbf{r},t)$ is obtained by a 2D demodulation of the movies (see~\cite{Cobelli}). Eventually we compute the velocity field $v(\mathbf{r},t)=\partial\zeta/\partial t$.

\begin{figure}[!htb]
\centering
\includegraphics[width=8.5cm]{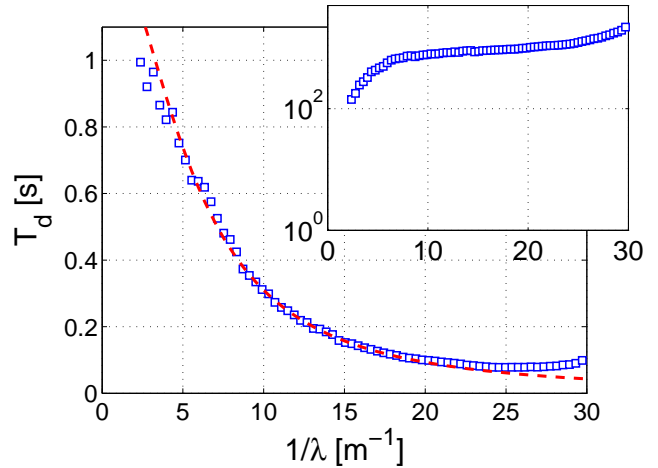}
\caption{(Color online) Squares: Measured dissipative time $T_d$ versus $k/2\pi=1/\lambda$. Dashed line: lorentzian fit $T_d=(0.73 + 0.025(k/2\pi)^2)^{-1}$ used in the following. Insert: $T_d\omega$ versus $k/2\pi$ (semilog scale).}
\label{dec}
\end{figure}
Our previous work show that the scale separation $T\ll T_d$ is verified in wave turbulence in an elastic plate~\cite{Miquel}. We studied the decay of energy of the wave turbulent field after stopping the forcing. A first period of decay is associated to the nonlinear cascade and is followed by a final exponential decay. We extract from this exponential part the dissipative characteristic time $T_d$ shown in fig.~\ref{dec}. The ratio of $T_d$ and the period of the wave is over 100 in the accessible range of wavelengths. This scale separation enables the existence of an intermediate nonlinear time scale.

In order to check the above scale separation of the nonlinear time scale in respect to $T$ and $T_d$, we perform an analysis of the dynamics of wave packets through a wavelet decomposition. The signal processing yielding the estimation of the nonlinear time scale is described in part~\ref{SP}. The scale separation is discussed in section~\ref{disc}.

%%%%%%%%%%%%%%
\section{Description of the wavelet analysis}
\label{SP}
%%%%%%%%%%%%%%

\subsection{Definition of the nonlinear time $T_{NL}$}

In the framework of the WTT, the amplitude of the wave is supposed to be slowly modulated by the nonlinear coupling between the waves. Let us write $\tilde v_{\mathbf k}(t)=\dfrac{1}{2\pi}\int v(\mathbf r,t)e^{j\mathbf k\cdot \mathbf r}d^2\mathbf r$ the spatial Fourier transform of the velocity field. The WTT assumes that $\tilde v_{\mathbf k}(t)$ can be written as 
\begin{equation}
\tilde v_{\mathbf k}(t)=a_{\mathbf k}(t)e^{-j\omega_{\mathbf k}t}
\end{equation}
where $a_{\mathbf k}(t)$ is the  amplitude which evolves slowly in time compared to the period of the wave. We define the nonlinear time scale as 
\begin{equation}
T_{NL}=\int_0^\infty\frac{|\mathcal D_{\mathbf k}(\tau)|}{|\mathcal D_{\mathbf k}(0)|}d\tau
\end{equation}
where $\mathcal D_{\mathbf k}(\tau)=\langle a_{\mathbf k}(t)a^*_{\mathbf k}(t+\tau)\rangle$ is the temporal correlation of the slow modulation of the wave. The average $\langle \, \rangle$ is a statistical average on realizations of the turbulent field and is estimated as a time average in our case (the system is statistically stationnary).
Unfortunately $\mathcal{D}_\mathbf{k}$ cannot be estimated directly from the experimental data because the velocity field is measured only on a fraction of the plate. A wavelet decomposition is performed to gain access to the slow modulation correlation function $\mathcal{D}_\mathbf{k}$.

\subsection{The wavelet family}
%%%%%%%%%%%%%

\begin{figure}[!htb]
\centering
\includegraphics[width=8.5cm]{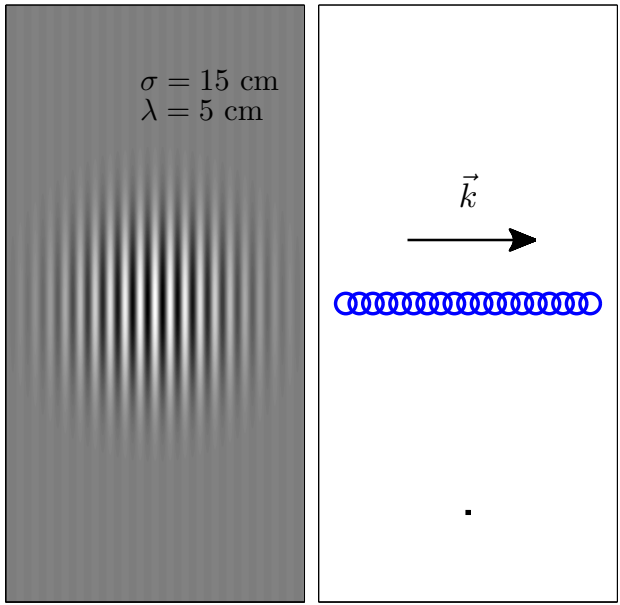}
\caption{(Color online) Schematics of the plate (figured by the vertical rectangular box) of size $2\times1$~m$^2$. Left: Example of the Gabor wavelets used in the analysis. The width of the Gaussian envelop is $\sigma=15$~cm and $k/2\pi=15\ \mathrm{m}^{-1}$. Right: collection of  positions of the center of the wavelets $\mathbf{R}$ used in our analysis. $\mathbf{k}$ is chosen horizontal. The bottom black dot corresponds to the position of the electromagnetic shaker.}
\label{exw}
\end{figure}

In order to extract the dynamics of a wavetrain at some wavevector $\mathbf k$ and position $\mathbf R$ we use the continuous family of 2D Gabor  wavelets:
\begin{equation}
G_{\mathbf k,\mathbf R}(\mathbf r)=\frac{1}{\sqrt{\pi\sigma^2}}\exp\left(-\frac{(\mathbf r-\mathbf R)^2}{2\sigma^2}\right)\exp(j\mathbf k\cdot \mathbf r)\, .
\end{equation}
Such a wavelet is a complex exponential of wavevector $\mathbf k$ modulated by a Gaussian envelop of width $\sigma$ and centered in $\mathbf R$. The choice of the width of the envelop results from a balance between the locality in space and in scale. In all the following we use a constant width $\sigma= 14.4$~cm for convenience (rather than fixing the product $\sigma k$ for instance). An illustration of the wavelet is shown in fig.~\ref{exw}. We define the wavelet coefficients as
\begin{equation}
\tilde{w}_{\mathbf k,\mathbf R}(t)=\int v(\mathbf r,t)G^*_{\mathbf k,\mathbf R}(\mathbf r)d^2\mathbf r
\end{equation}
where $^*$ is the complex conjugation. The wavelet coefficient is the projection of the turbulent mix of waves at time $t$ on a Gaussian wave packet at position $\mathbf R$ and wavevector $\mathbf k$.

\subsection{Correlations of wavelet coefficients}
%%%%%%%%%%%%%%%%%%%%%%%%

In order to extract the nonlinear dynamics of the turbulent field, we investigate the time correlations of the wavelet coefficients:
\begin{eqnarray}
\mathcal{A}_{\mathbf k,\mathbf R,\mathbf R'}(\tau)&=&\left\langle \tilde{w}_{\mathbf k,\mathbf R}(t)\tilde{w}^*_{\mathbf k,\mathbf R'}(t+\tau)\right\rangle\\
\mathcal{C}_{\mathbf k,\mathbf R,\mathbf R'}(\tau)&=&\left\langle \tilde{w}_{\mathbf k,\mathbf R}(t)\tilde{w}^*_{-\mathbf k,\mathbf R'}(t+\tau)\right\rangle
\end{eqnarray}
$\mathcal A$ is the correlation of the amplitude of a wave packet propagating along the same $\mathbf k$ and between the positions $\mathbf R$ and $\mathbf R'$ and at times separated by $\tau$. $\mathcal C$ is a similar correlation but for wave packets propagating along opposite wavevectors. Here we study only correlations between positions on a line parallel to the horizontal wavevector $\mathbf k$ as shown in fig.~\ref{exw}. The same correlations are computed for a plate at rest as well in order to get an estimate of the noise contribution. Assuming that the noise and the signal are independent, the contribution of the noise can be subtracted from the direct estimation of $\mathcal A$ so that to increase the dynamics of the measurement and in particular to remove contributions due to the flickering and the scanning of the videoprojector (acting at very low $k$ and at 60~Hz and its harmonics).

\subsection{Extraction of the nonlinear dynamics}
%%%%%%%%%%%%%%%%%%%%%%%%%

In this subsection, we describe how the slow modulation correlation $\mathcal{D}$ can be computed from the measured coefficients $\mathcal{A}$ and $\mathcal{C}$.
The correlation $\mathcal A$ can be rewritten in Fourier space as
\begin{equation}
\mathcal{A}_{\mathbf k,\mathbf R,\mathbf R'}(\tau)=\int \langle\tilde{v}_{\mathbf q}(t)\tilde{v}^*_{\mathbf q'}(t+\tau)\rangle\tilde G^*_{\mathbf k,\mathbf R}(\mathbf q)\tilde G_{\mathbf k,\mathbf R'}(\mathbf q')d^2\mathbf q d^2\mathbf q'
\end{equation}
where $\tilde G_{\mathbf k,\mathbf R}(\mathbf q)$ is the spatial Fourier transform of the wavelet $G_{\mathbf k,\mathbf R}(\mathbf r)$.

In the framework of the WTT, two Fourier modes of the turbulent field are supposed to be uncorrelated at different wavevectors (correlations appear for at least 3 waves). It has been checked in this experiment for 2-wave equal time correlations in~\cite{epjb}.  Using this hypothesis, one gets: 
\begin{eqnarray}
\langle\tilde{v}_{\mathbf q}(t)\tilde{v}^*_{\mathbf q'}(t+\tau)\rangle&=&\langle\tilde{v}_{\mathbf q}(t)\tilde{v}^*_{\mathbf q}(t+\tau)\rangle\delta(\mathbf q-\mathbf q')\\
&=& \delta(\mathbf q-\mathbf q')\mathcal D_{\mathbf q}(\tau) e^{j\omega_{\mathbf q}\tau}
\end{eqnarray}

The correlation can thus be rewritten as 
\begin{equation}
\mathcal{A}_{\mathbf k,\mathbf R,\mathbf R'}(\tau)=\int \tilde G^*_{\mathbf k,\mathbf R}(\mathbf q)\tilde G_{\mathbf k,\mathbf R'}(\mathbf q) \mathcal D_{\mathbf q}(\tau)e^{j\omega_{\mathbf q}\tau}d^2\mathbf q\, .
\label{eqA}
\end{equation}

We assume in the following that $\mathcal D_{\mathbf q}(\tau)$ is varying more slowly as a function of $\mathbf q$ than the Fourier transform $G_{\mathbf k,\mathbf R}(\mathbf q)$. By doing this assumption, we suppose that the extension of the wavelet (in real space) is narrower than the spatial coherence of the wave. As the wavelet has a Gaussian envelop, $G_{\mathbf k,\mathbf R}(\mathbf q)$ is also Gaussian centered on $\mathbf q=\mathbf k$. With the above assumption one can neglect the variation of $\mathcal D$ in the equation~(\ref{eqA}) so that:
\begin{equation}
\mathcal{A}_{\mathbf k,\mathbf R,\mathbf R'}(\tau)\approx  \mathcal D_{\mathbf k}(\tau)\int \tilde G^*_{\mathbf k,\mathbf R}(\mathbf q)\tilde G_{\mathbf k,\mathbf R'}(\mathbf q)e^{j\omega_{\mathbf q}\tau}d^2\mathbf q\, .
\end{equation}
The integral in this equation can be calculated using the dispersion relation of the waves in the elastic plate $\omega = c k^2$ and yields
\begin{equation}
\mathcal{A}_{\mathbf k,\mathbf R,\mathbf R'}(\tau)=\frac{\mathcal D_{\mathbf k}(\tau)}{\sqrt{1+\frac{c^2\tau^2}{\sigma^4}}}\exp\left(-\frac{\left(\Delta \mathbf R-\mathbf v_g(\mathbf k)\tau\right)^2}{4\sigma^2(1+\frac{c^2\tau^2}{\sigma^4})}\right)e^{j\Phi}
\label{analytical_prediction}
\end{equation}
where $\Delta \mathbf R=\mathbf R'-\mathbf R$, $\mathbf v_g(\mathbf k)=2c\mathbf k$ is the group velocity and $\Phi=\omega\tau+\phi(\omega,\tau)$ where $\phi$ is a phase factor.

The correlation $\mathcal A$ of the wavelet coefficients contains the variations of the correlation of the modulation of the wave $\mathcal D$ but altered by the factor 
\begin{equation}
\frac{1}{\sqrt{1+\frac{c^2\tau^2}{\sigma^4}}}\exp\left(-\frac{\left(\Delta \mathbf R-\mathbf v_g(\mathbf k)\tau\right)^2}{4\sigma^2(1+\frac{c^2\tau^2}{\sigma^4})}\right)e^{j\Phi}
\end{equation}
which is due to the dispersion of the wavelet during its propagation from $\mathbf R$ to $\mathbf R'$ at the group velocity. The prefactor $\left(1+\frac{c^2\tau^2}{\sigma^4}\right)^{-1/2}$ can be easily compensated. We study rather the compensated correlations in the following: $\mathcal A^c=\sqrt{1+\frac{c^2\tau^2}{\sigma^4}}\mathcal A$ and $\mathcal C^c=\sqrt{1+\frac{c^2\tau^2}{\sigma^4}}\mathcal C$.  The modulus of the compensated coefficient $\mathcal{A}^c$ is directly linked to $\mathcal{D}$:
\begin{equation}
\left |\mathcal{A}^c_{\mathbf k,\mathbf R,\mathbf R'}(\tau)\right |=|\mathcal D_{\mathbf k}(\tau)|\exp\left(-\frac{\left(\Delta \mathbf R-\mathbf v_g(\mathbf k)\tau\right)^2}{4\sigma^2(1+\frac{c^2\tau^2}{\sigma^4})}\right)
\label{model_Ac}
\end{equation}
Wave packet propagation over a distance $\Delta \mathbf R$ at the group velocity $\mathbf v_g$ account for the Gaussian additionnal term. Note that in our configuration, this extra term equals 1 for $\tau=\epsilon\frac{|\Delta \mathbf R|}{|\mathbf v_g(\mathbf k)|}$ with $\epsilon$ being the sign of $\Delta \mathbf R\cdot\mathbf v_g$. This time is the flight time of the wave packet at the group velocity.
Thus by scanning a collection of positions $\mathbf R, \mathbf R'$ the magnitude of the correlation $|\mathcal D|$ can be reconstructed at all time lags $\tau$ and the nonlinear time scale $T_{NL}$ can be computed. We emphasize that the method has been checked to be robust: the result does not depend on the choice of the wavelet envelope (although we restrain our study to Gaussian wavelets in this paper for tractability sake) nor on the choice of the width of the wavelet $\sigma$.

\subsection{Experimental results and rebounds}
%%%%%%%%%%%%%%%%%%%%%%%%
\begin{figure}[!htb]
\centering
\includegraphics[width=8.5cm]{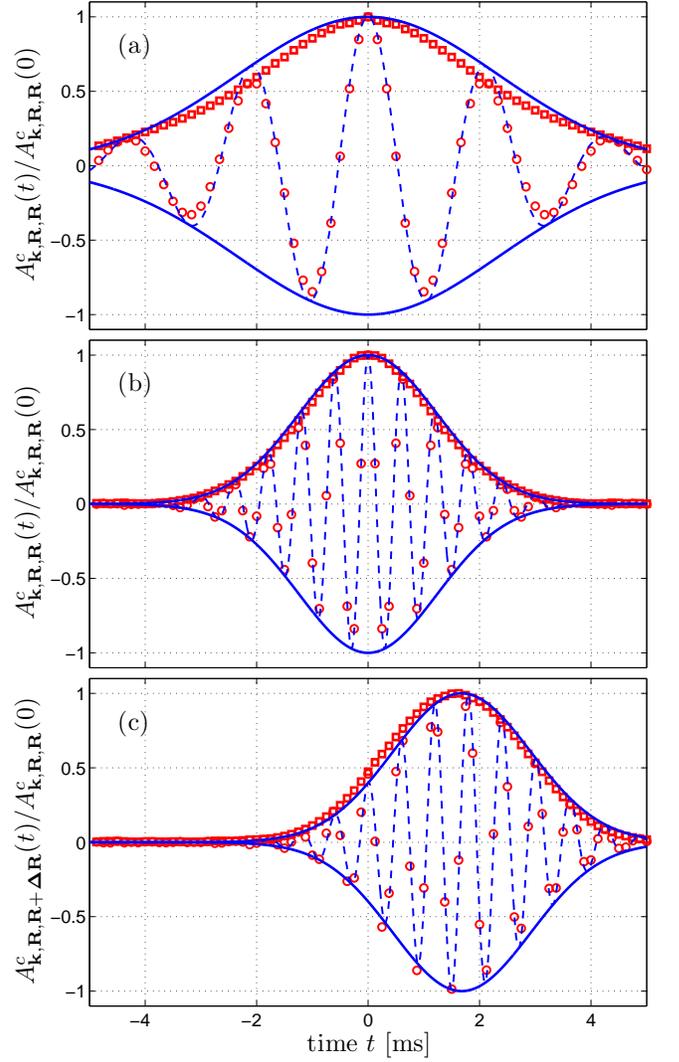}
\caption{(Color online) Modulus $|\mathcal{A}^c(t)|$ (red squares: experimental results; blue solid lines: analytical predictions) and real part $|\mathcal{A}^c(t)|$ (red circles: experimental results; blue dashed lines: analytical predictions) of the compensated correlation functions $\mathcal{A}^c_{\mathbf{k,R,R+\Delta R}}(t)$. No adjustable parameter is used. (a) $k/(2\pi)=10.6\ \mathrm{m}^{-1}$, $\mathbf{\Delta R}=0$; (b) $k/(2\pi)=20.2\ \mathrm{m}^{-1}$, $\mathbf{\Delta R}=0$; (c) $k/(2\pi)=20.2\ \mathrm{m}^{-1}$, $\mathbf{\Delta R}=0.27\ \mathrm{m}$. }
\label{gausscor}
\end{figure} 

\begin{figure}[!htb]
\centering
\includegraphics[width=8.5cm]{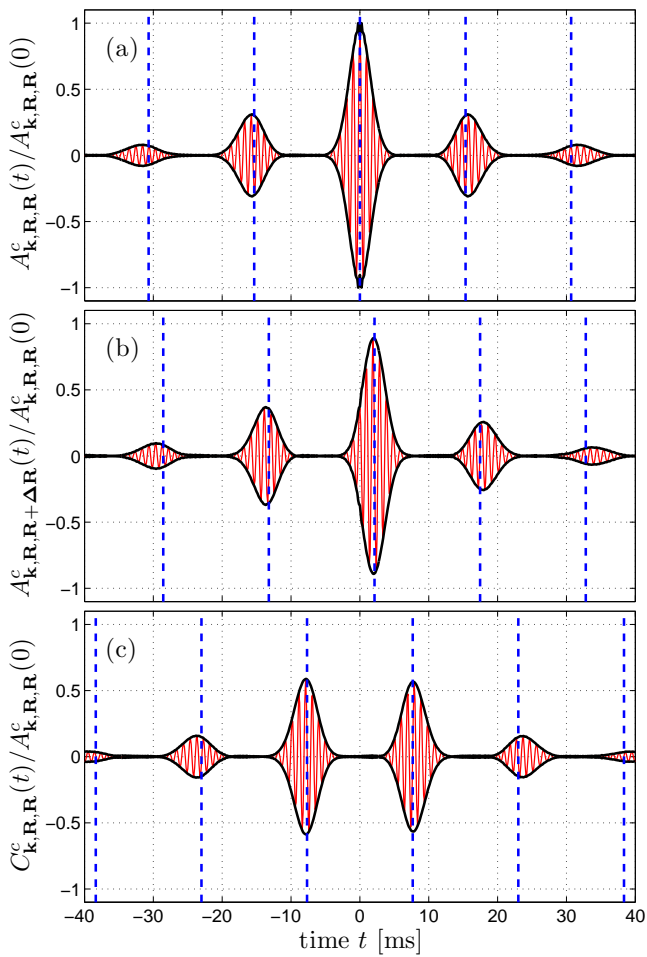}
\caption{(Color online) Example of correlations at $k=16.0$~m$^{-1}$. The real part is shown as the oscillating thin red line and the modulus as the thick black line. (a) $\mathcal{A}^c_{\mathbf k,\mathbf R,\mathbf R}(\tau)$ for $\mathbf R$ at the center of the plate. The vertical dashed lines show the times $2nL/v_g$ for $n=-2$ to $2$. (b) $\mathcal{A}^c_{\mathbf k,\mathbf R,\mathbf R'}(\tau)$ for $\Delta R=0.27$~m. The vertical dashed lines show the times $( 2nL + \Delta R)/v_g$ for $n=-2$ to $2$. (c) $\mathcal{C}^c_{\mathbf k,\mathbf R,\mathbf R}(\tau)$ for $\mathbf R$ at the center of the plate. The vertical dashed lines show the times $(2n+1)L/v_g$ for $n=-3$ to $2$. 
In the three cases, the correlations have been normalized by $\mathcal{A}_{\mathbf k,\mathbf R,\mathbf R}(0)$.}
\label{excor}
 \end{figure}
In this subsection we present a comparison between the experimental data and eq.~\eqref{model_Ac} predicted by our model: examples of compensated correlation functions $\mathcal{A}^c_{\mathbf k,\mathbf R,\mathbf R'}(\tau)$ and $\mathcal{C}^c_{\mathbf k,\mathbf R,\mathbf R'}(\tau)$ are shown in fig ~\ref{gausscor} and fig.~\ref{excor}. Figure~\ref{gausscor} compares measured correlations for short times with the expression given in eq~\eqref{model_Ac}. The modulus of the correlations at the same positions $\mathbf R=\mathbf R'$ (fig~\ref{gausscor}(a) and (b)) exhibit a maximum for $\tau=0$. As $\tau$ grows, the wavelet propagates away from its original position and this modulus decreases to 0 in the Gaussian fashion predicted by equation~\eqref{analytical_prediction}. The real part oscillates as predicted by the term $\exp\left(j\left[\omega t+\phi\right]\right)$. As expected, a wave at a larger wavenumber propagates further and oscillates faster than for smaller wavevectors. An example of correlation for different positions is displayed on figure~\ref{gausscor}(c). The maximum is delayed due to the flight time. Measurements shows a very good agreement with predictions, although a small departure from the expected Gaussian envelop is observed for small wavevectors / large wavelengths (figure~\ref{gausscor}(a)). Indeed, the second statement of our derivation is not met for small wavenumbers: the wavelength and the size of the wavelets are comparable.
 
The measured correlation coefficients are plotted for larger time lags on figure~\ref{excor}. The autocorrelation coefficients $\mathcal{A}$ (fig~\ref{excor}(a,b)) exhibit some secondary maxima. They correspond to a wave packet that has bounced on the borders of the plate. After one rebound, the wave packet propagates back with the opposite wavevector $-\mathbf k$. Thus no correlation is observed until it bounces a second time and propagates again with its original wavevector. Correlation reappear as it comes back to its initial position $\mathbf R$. These reflexions account for repeated maxima with a delay $T_\mathrm{fly}=2L/v_g$ corresponding to the wave packet round trip accross the plate. These rebounds can be used to extend the range of accessible values of the time lag for the reconstruction of $\mathcal D$. The magnitude of the reflexion coefficient at the border is expected to be very close to unity for a free border moving in air (whose density is about 4 orders of magnitude lower than steel). The relative amplitude of the maxima is decaying with the time lag because of the decoherence of the wave as it propagates.

Figure~\ref{excor}(c) shows $\mathcal C^c$ for $\mathbf R=\mathbf R'$. As expected, the correlation is zero at small $\tau$ because the wave packet has not enough time to propagate to the border of the plate and back to the orginal position. A maximum of the modulus of the correlation is observed for a time lag $\tau=L/v_g$. This corresponds to the time needed for a wave packet to travel to the border, bounce on it and come back to its original position counterpropagating. Secondary maxima are also observed with a period $T_\mathrm{fly}$. By collecting all values of all maxima of $\mathcal{A}^c$ and $\mathcal{C}^c$ one can build the modulus of $\mathcal D$ for quite long time lags (until the signal falls under the noise level).

\begin{figure}[!htb]
\centering
\includegraphics[width=8.5cm]{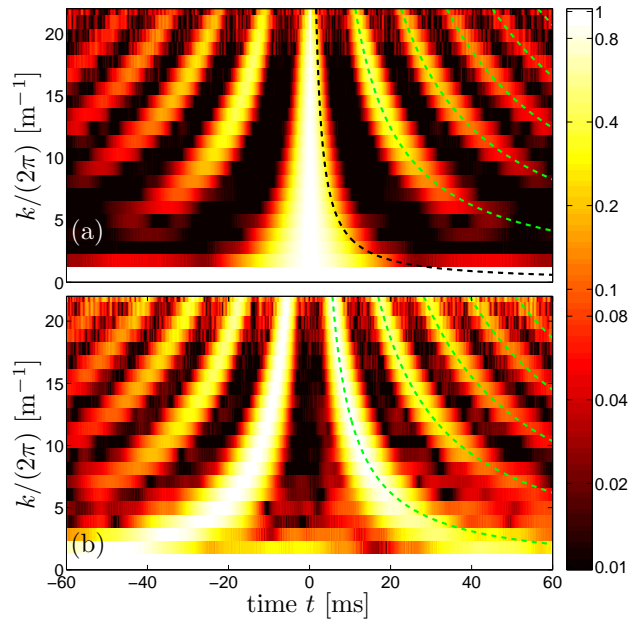}
\caption{(Color online) Modulus of the correlations at the same position $\mathbf R=\mathbf R'$ as a function of time and wavevector. Colors are log coded. (a) Autocorrelations $|\mathcal A^c|$. Green (light gray) dashed lines: flight times for $2n$ rebounds $t_n^\mathrm{fly}=2nL/(v_g(k))$; black dashed line: flight time over the width of the wave packet (b) Crosscorrelations $|\mathcal C^c|$. Green (light gray) dashed line: flight times for $2n-1$ rebounds $t_n^\mathrm{fly}=(2n-1)L/(v_g(k))$. At each value of $k$, the correlations have been normalized by $\mathcal{A}^c_{\mathbf k,\mathbf R,\mathbf R}(0)$ so that the modulus of the correlation is between 0 and 1.}
\label{ods}
\end{figure}
 Figure~\ref{ods} presents in its upper part the correlation $\mathcal{A}^c_{\mathbf k,\mathbf R,\mathbf R}(\tau)$ as a function of both $k$ and $\tau$ and $\mathcal{C}^c_{\mathbf k,\mathbf R,\mathbf R}(\tau)$ in the lower part. The dashed lines correspond to flight times over $2nL$ for $\mathcal A$ or $(2n+1)L$ for $\mathcal C$ at the group velocity. The local maxima of correlation follow these lines: this validates our interpretation of the secondary maxima as due to the rebounds. 
 
\begin{figure}[!htb]
\centering
\includegraphics[width=8.5cm]{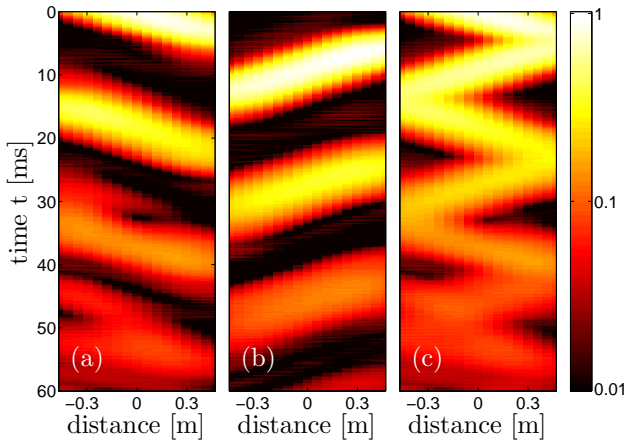}
\caption{(Color online) (a) Color plot of $\left|\mathcal{A}^c_{\mathbf{k,R,R'}}(t)\right| / \left|\mathcal{A}^c_{\mathbf{k,R,R}}(0)\right|$ versus time $t$ and position $\mathbf{R'}$. Wavenumber $k/2\pi=13\ \mathrm{m}^{-1}$; the position $\mathbf{R}$ is at center of the plate. Colors are log-coded. (b) Crosscorrelation $\left|\mathcal{C}^c_{\mathbf{k,R,R'}}(t)\right| / \left|\mathcal{A}^c_{\mathbf{k,R,R}}(0)\right|$ (c) Combinaison of the two functions $\mathcal A^c$ and $\mathcal C^c$: $\mathcal{T}=\sqrt{\left|\mathcal{A}^c\right|^2+\left|\mathcal{C}^c\right|^2}$.}
\label{zz1} 
\end{figure}
A picture of the travel of the wave packet can be build by plotting the modulus of the correlation in a space time representation as in fig.~\ref{zz1}. Figure~\ref{zz1}(a) and (b) show the modulus of $\mathcal A^c$ and $\mathcal C^c$ respectively. Figure~\ref{zz1}(a) displays only the wave packet traveling from left to right whereas fig.~\ref{zz1}(b) shows only the reverse motion. The trajectories appear slightly deformed in a S shape due to the fact that close to the border of the plate, the wave packet is partially reflected in the other direction and the wavelet is truncated by the border. The two functions can be combined to show the full trajectory in fig.~\ref{zz1}(c). The correlation is seen to decay in magnitude as time goes by along the trajectory corresponding to the loss of temporal coherence of the wave packet due to the energy exchanges with the other waves by the nonlinear coupling. 

\begin{figure}[!htb]
\centering
\includegraphics[width=8.5cm]{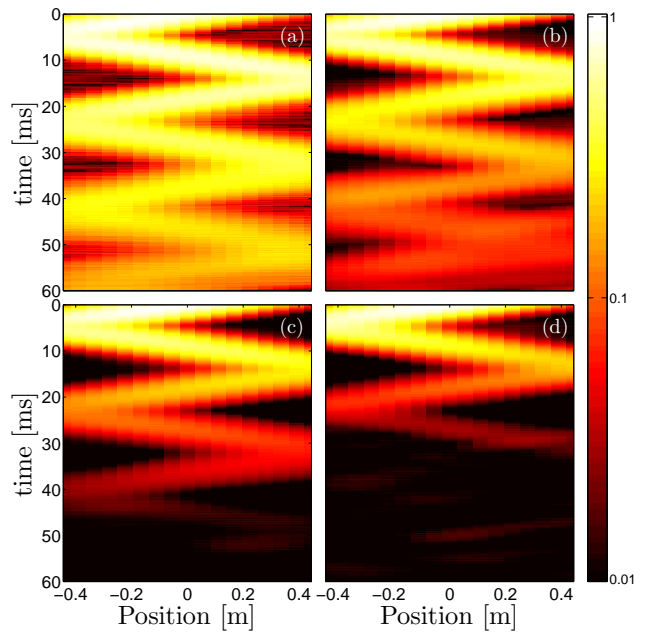}
\caption{(Color online) Color plot of $\mathcal T$ (see previous figure) versus time $t$ and positions $\mathbf{R'}$. Wavenumber $k/2\pi=13\ \mathrm{m}^{-1}$; the position $\mathbf{R}$ is the center of the plate. Colors are log-coded. {The forcing power is $P_0$, $7P_0$, $27P_0$ and $59P_0$ (from (a) to (d))}}
\label{zz2} 
\end{figure}
Figure~\ref{zz2} shows the space-time representation of the decay of the correlation along the trajectory of the wave packet for 4 different values of the forcing. As the forcing is increased, the number of rebounds that can be observed is also decaying: 6 bounces can be seen in the 60~ms time window at the lowest forcing while only three are visible at the strongest forcing intensity. The stronger the forcing, the higher the nonlinearity, the faster the energy transfer and thus the smaller the nonlinear time.
 
%%%%%%%%%%%%%%%%%%%% 
\subsection{Nonlinear time measurement}\label{TNL}
%%%%%%%%%%%%%%%%%%%%
 \begin{figure}[!htb]
\centering
\includegraphics[width=8.5cm]{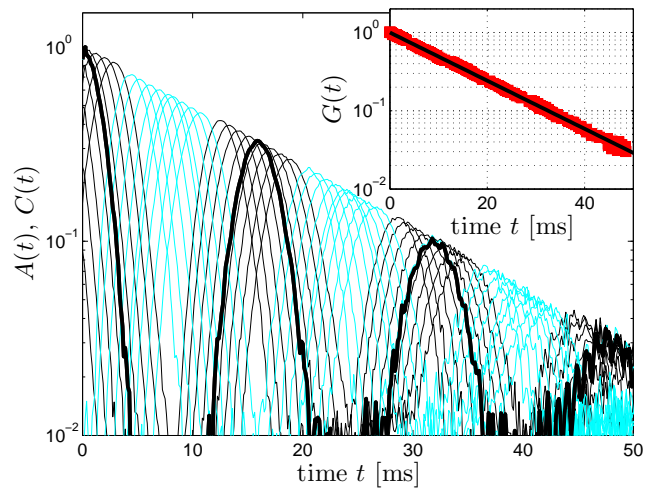}
\caption{(Color online) Superimposition of the modulii $|\mathcal{A}^c_{\mathbf k,\mathbf R,\mathbf R'}(\tau)|$ (normalized by $|\mathcal{A}^c_{\mathbf k,\mathbf R,\mathbf R}(0)|$ at the center) at a given $k$ and for a collection of positions $\mathbf R'$ (excluding positions too close to the border for which the wavelet can be truncated). $\mathbf R$ is chosen at the center of the plate. The thick line corresponds to $\mathbf R=\mathbf R'$. Black thin lines: $\mathcal{A}^c_{\mathbf k,\mathbf R,\mathbf R'}(\tau)$; cyan (light gray) thin lines: $\mathcal{C}^c_{\mathbf k,\mathbf R,\mathbf R'}(\tau)$.
 Insert: squares correspond to the value of the local maximum. The line is an exponential fit of the decay.}
\label{tnl}
\end{figure}
In order to rebuild the modulus of the correlation $\mathcal D_{\mathbf k}(\tau)$, one should collect the values of the local maxima of the correlations $\mathcal A^c$ and $\mathcal C^c$ for various pairs of positions $\mathbf R,\mathbf R'$. An example of  collection of all such correlation curves is shown in fig.~\ref{tnl} for given $\mathbf k$ and forcing intensity. The maxima are extracted by a local Gaussian fit and shown in the insert. The decay of the maxima with the time lag is seen to be exponential so that  
\begin{equation}
\frac{|\mathcal D_{\mathbf k}(\tau)|}{|\mathcal D_{\mathbf k}(0)|}\propto \exp -\frac{\tau}{T_{NL}(k)}
\end{equation}
which permits a direct extraction of the nonlinear time scale.

%%%%%%%%%%%%
\section{Discussion}\label{disc}
%%%%%%%%%%%%

\subsection{The nonlinear time scale}
%%%%%%%%%%%%%%%%%%%

\begin{figure}[!htb]
\centering
\includegraphics[width=8.5cm]{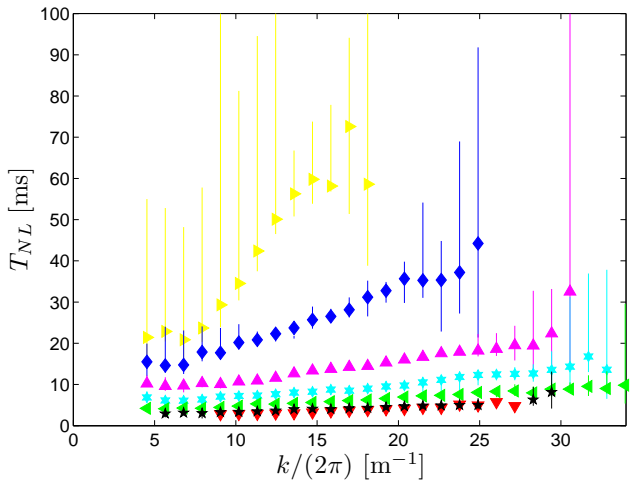}

\caption{(Color online) Evolution of the nonlinear time scale as a function of the wavenumber for different forcing. Symbols are experimental data for different values of the forcing intensity (intensity increasing downward): $P_0$, $7P_0$, $27P_0$, $59P_0$, $100P_0$, $149P_0$, and $207P_0$. This colorchart for injected powers is used through this section for other figures.}
\label{tnlexp} 
\end{figure}
The evolution of the nonlinear time scale with the wavenumber and the forcing intensity is shown in fig.~\ref{tnlexp}. The estimation of $T_{NL}$ was not performed for $k/2\pi$ less than 5 as the wavelength is too large to fit in the Gaussian modulation of the wavelet (the wavelength is comparable to $\sigma$). At large $k$, the estimation is performed until the signal over noise ratio gets too degraded to allow any computation of the correlations. Thus the highest wavenumber that is reached depends on the amount of data gathered at the considered value of the forcing as well as on the frame rate of the camera (the highest the frame rate, the lowest the amount of light, the highest the noise level). The nonlinear time may be considered as a measure of nonlinearities: in this respect, we expect this time to diverge for linear problems and to decrease as the nonlinearities grows. Figure~\ref{tnlexp} shows that $T_{NL}$ is indeed a growing function of $k$ and a decreasing function of the forcing $P$. The evolution of $T_{NL}$ with $k$ is affine. 

\begin{figure}[!htb]
\centering
\includegraphics[width=8.5cm]{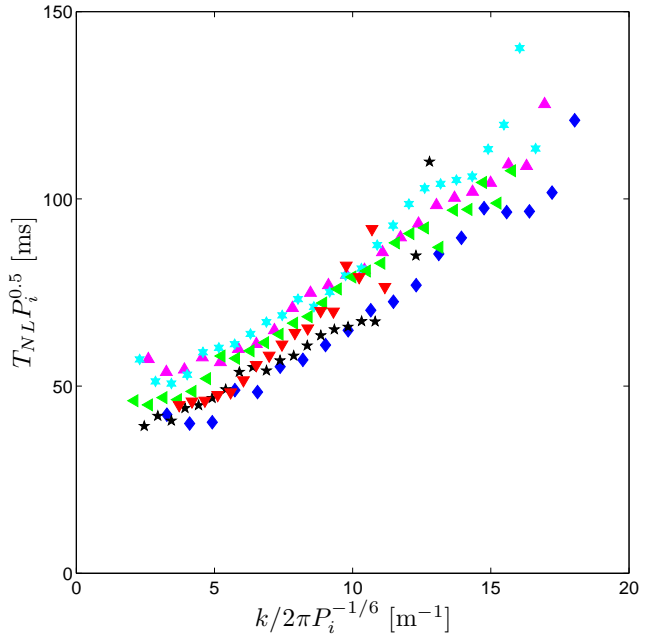}
\caption{(Color online) Rescaled nonlinear time $T_{NL}P^{0.5}$ as a function of the rescaled wavevector $kP^{-1/6}$ for different injected power $P$.}
\label{rescaling}
\end{figure}
A cutoff frequency was previously measured from the wave spectrum~\cite{Mordant}: $f^*\propto P^{1/3}$ which translates through the dispersion relation into $k^*\propto P^{1/6}$. This cutoff frequency corresponds to the transition to a dissipative regime. Here we look for the following rescaling of the nonlinear time scale:
\begin{equation}
T_{NL}(k)\propto P^{-\alpha} f\left(k P^{-1/6}\right)
\label{eq_tnl_scal}
\end{equation}
The curves for different forcing collapse on a master curve when using $\alpha=0.5$ (fig.~\ref{rescaling}). This scaling is similar to that previously observed for the power spectrum~\cite{Boudaoud, Mordant}.
In the framework of the WTT, one would expect that $T_{NL}$ scales as $P^{-1/3}$ for 4-wave resonance as predicted by the theory~\cite{During} or as $P^{-1/2}$ for 3-wave resonance. Here, the rescaled time is affine with the rescaled wavenumber so that a pure scaling is not observed. The nonlinear time follows a scaling law
\begin{equation}
T_{NL}=\beta P^{-1/2}+\gamma k P^{-2/3}
\end{equation} 
with constant $\beta,\gamma$.

\subsection{Time scale separation}
%%%%%%%%%%%%%%%%%

\begin{figure}[!htb]
\centering
\includegraphics[width=8.5cm]{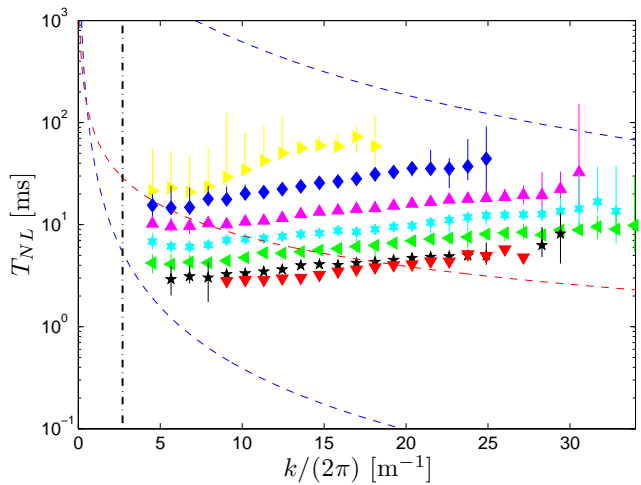}
\caption{(Color online) Comparison of the nonlinear time scale with other relevant time scales. Symbols are experimental data for $T_{NL}$ at different values of the forcing intensity. Upper dashed line: twice the dissipative time $T_d$ measured by the decay of energy in the unforced case (see fig.\ref{dec}). Middle dashed line: $2/\delta\omega_{sep}=L/(\pi k c)$. Lower dashed line: $1/\omega$. The forcing is operated at 30~Hz and is shown with the vertical dash-dotted line.}
\label{sep} 
\end{figure}

The experimental time scales are compared with the dissipative time scales and the period of the wave in fig.~\ref{sep}. The figure displays rather twice the time $T_d$ estimated from the decay of the energy in the decline case as the energy is quadratic in the amplitude of the wave thus the decay time of the amplitude is twice that of the energy. The figure shows $1/\omega$ rather than the period $2\pi/\omega$ for the following reason: the correlation of the magnitude of the wave is observed to be exponentially decaying. Its Fourier transform follows thus a Lorentzian shape $\propto 1/\left(\omega^2+\frac{1}{T_{NL}^2}\right)$. Thus $T_{NL}$ should be compared directly to $1/\omega$. 

\begin{figure}[!htb]
\centering
\includegraphics[width=8.5cm]{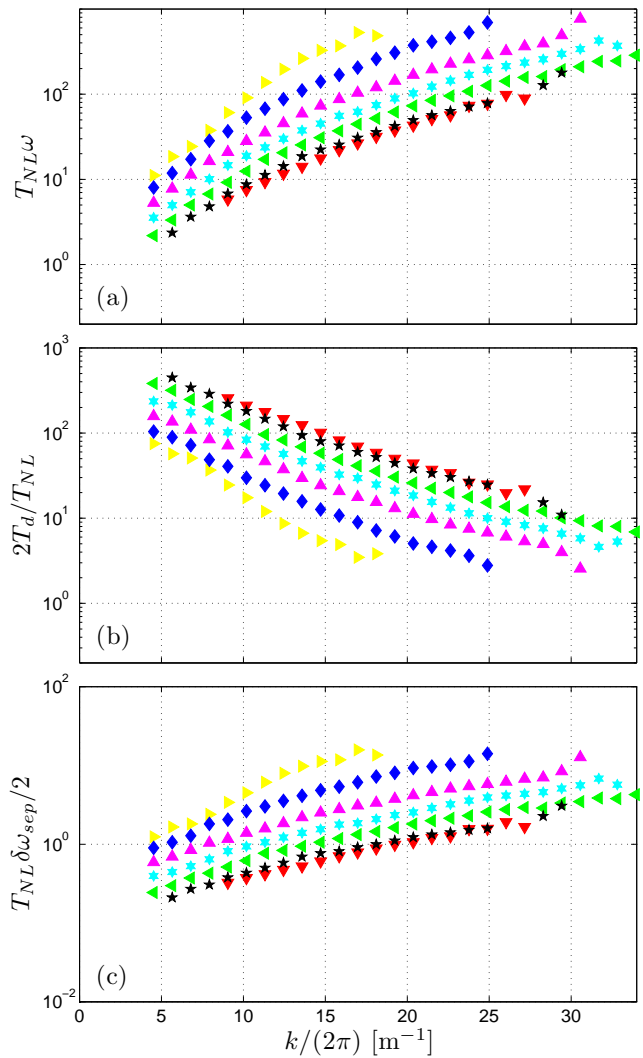}
\caption{(Color online) Ratios between different characteristic times as a function of $k$ for various values of the injected power: (a) $\omega T_{NL}$; (b) $2T_d/T_{NL}$; (c) $T_{NL}\delta \omega_{sep}/2$ }
\label{ratio}
\end{figure}

The product $T_{NL}\omega$ is shown in figure~\ref{ratio}(a). The nonlinear time scale is seen to be over an order of magnitude above $\omega$ for most of the accessible wavenumber range so that the scale separation $\omega T_{NL}\gg1$ is verified. At the highest forcing intensity the ratio is only a factor 2 at the forcing scale suggesting that the hypothesis of weak nonlinearity may be disputable at this scale. 

The ratio $2T_d/T_{NL}$ is displayed in figure~\ref{ratio}(b). The nonlinear time scale is over two orders of magnitude shorter than the dissipative time at the forcing scale. The ratio is always larger than 3 at the highest wavenumber that can be analyzed (which is always above the observed dissipative cutoff). In the framework of the Weak Turbulence Theory one would expect the conservative energy cascade to operate until $T_{NL}\simeq T_d$ and to be followed by a dissipative regime. Here it appears that the systems reaches the dissipative regime at wavenumbers for which the nonlinear time are still larger than the dissipative times. The cascade is stopped by another phenomenon before reaching the expected dissipative region. This observation is attributed to finite size effects in the following section.

\subsection{Finite size effects and transition to frozen turbulence}
%%%%%%%%%%%%%%%

Finite size effects can be neglected when the half separation of the modes $\delta \omega_{sep}/2$ is larger than $1/T_{NL}$. The quantity $2/\delta \omega_{sep}$ is compared with $T_{NL}$ in fig.~\ref{sep} and the ratio is shown in figure~\ref{ratio}(c). At low forcing intensity, the product $T_{NL}\delta \omega_{sep}$ is always larger than one so that finite size effects are expected to be dominant over the whole interval of accessible wavenumbers. When increasing the forcing, the product gets smaller. At the strongest forcing the ratio is lower than one over the whole interval of $k$ so that a behavior close to WTT should be expected. At intermediate values of the forcing a crossover is seen from a regime were finite size effects should be negligible (at low $k$) to a regime dominated by finite size effects (at large $k$). The crossover region corresponds to laminated turbulence as described in \cite{Kartashova,Kartashova1,Lvov}. In this regime, the resonance conditions on frequencies $\omega_1+\omega_2=\omega_3+\omega_4$ (expected for 4-wave resonance~\cite{During}) are made less numerous due to the discretization of the modes. Nonlinear spectral widening allows for approximate resonances if the nonlinearities are strong enough so that a continuous regime can be observed. In the opposite case, only a few clusters of frequencies can be resonant and the cascade operates with a reduced efficiency. If the nonlinearity is too weak, the cascade is frozen when $T_{NL}\simeq\frac{2}{\delta \omega_{sep}}$ so that the energy cannot be transferred to higher wavenumbers and is ultimately dissipated. In this case, the energy cascade stops before reaching the dissipative regime expected from WTT (at which $T_{NL}\simeq T_d$).

\begin{figure}[!htb]
\centering
\includegraphics[width=8.5cm]{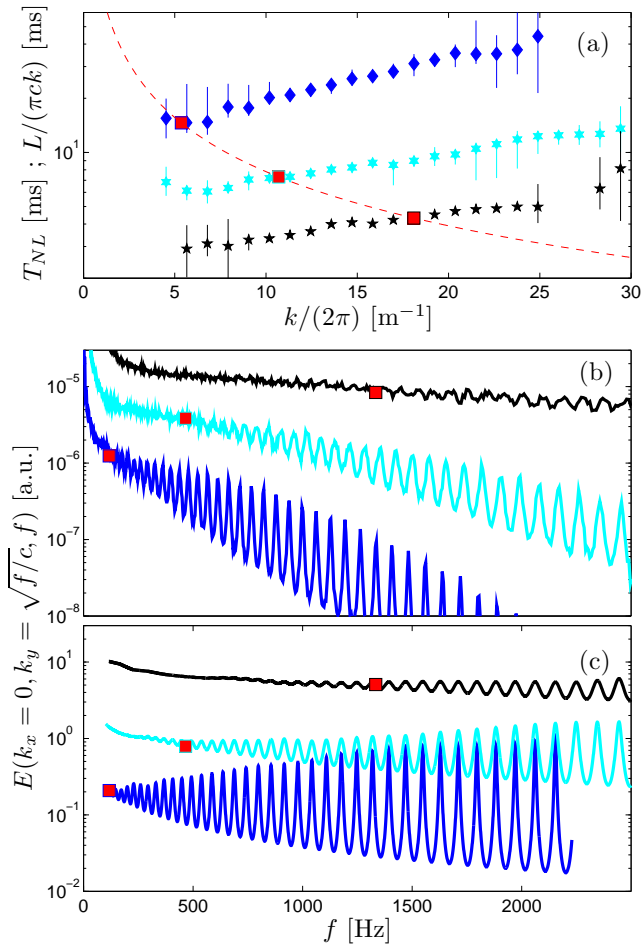}
\caption{(Color online) (a) Symbols: measured $T_{NL}$. Dashed line: $2/\delta \omega_{sep}$. {Red} filled squares: crossover wavenumber $k_c$ given by $T_{NL}(k_{c})=2/\delta\omega_{sep}(k_c)$. (b) Solid lines: Stationary power spectrum of the velocity extracted along the dispersion relation and for $k_x=0$ (see text). Forcing is increased from bottom to top. Plain squares: crossover frequency $f_c$ reported from (a). (c) Synthetic spectrum incorporating the nonlinear widening of the modes (see text). The width of the modes is $1/(2\pi T_{NL})$ interpolated as a function of frequency from (a). Curves are vertically shifted for clarity on fig (b,c).}
\label{kcross_discret}
\end{figure}
Figure~\ref{kcross_discret} analyses the experimental observation of the widening of the modes due to nonlinear effects. The cross-over wavevector $k_c$ defined as:
\begin{equation}
T_{NL}(k_c)=\frac{2}{\delta\omega_{sep}(k_c)}
\label{defkc}
\end{equation} is first determined on fig.~\ref{kcross_discret}(a) for different injected power. Using the dispersion relation, this crossover wavenumber is associated with a crossover frequency $f_c$:
\begin{equation}
f_c=ck_c^2/(2\pi)\label{deffc}
\end{equation} 

By performing Fourier transforms both in time and space, one computes the power spectrum density $E(\mathbf k,\omega)=\langle |v(\mathbf k,\omega)|^2\rangle$. 
Figure~\ref{kcross_discret}(b) shows a part of the power spectrum density $E(k_x=0, k_y=\sqrt{\omega/c},\omega)$, i.e. along the dispersion relation in a cut $k_x=0$. A transition can be observed between 2 distinct regimes: the spectrum is smooth and continuous at low frequencies whereas it displays some clearly separated peaks at high frequencies. This transition is well described by our criteria (eq~\eqref{deffc}): the crossover frequency extracted from figure~\ref{kcross_discret}(a) separates the two distinct regimes as shown in figure~\ref{kcross_discret}(b).  A toy model can be build to incorporate the nonlinear widening of discrete modes. The frequency of the modes are given by $f_n=c(n\pi /L)^2/(2\pi)$  and each mode is broadened following a Lorentzian variation of width $1/(2\pi T_{NL})$. A density of energy can be build in the following way:
\begin{equation}
\rho(f)\propto  \sum_{n}\frac{1}{\left[2\pi T_{NL}(f-f_n)\right]^2+1}
\label{mode}
\end{equation}  
This synthetic density of mode $\rho(f)$ is shown in fig.~\ref{kcross_discret}(c). The overall characteristics of fig.~\ref{kcross_discret}(b) are recovered (except for the general decay which is not taken into account in the model).

\subsection{Width of the dispersion relation}
%%%%%%%%%%%%%%%%%%%%%%

\begin{figure}
\centering
\includegraphics[width=\columnwidth]{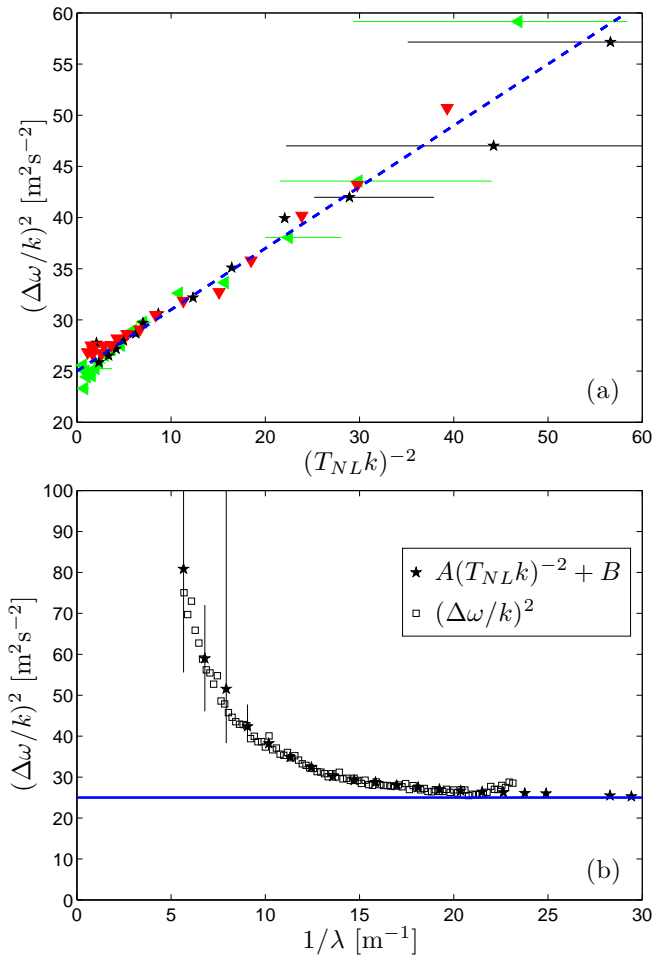}
 \caption{(Color online) (a)Circles: $(\Delta\omega/k)^2$ as a function of $(T_{NL}k)^{-2}$ for different injected powers. Dashed line (eyeguide): $0.6x+25$. (b)$A(T_{NL}k)^{-2}+B$ (circles) and $(\Delta\omega/k)^2$ (squares) plotted as a function of $k$ for one given injected power $P=149P_0$. $A=0.6$, $B=25$. Solid line: value of $B$. }
\label{large}
\end{figure}

Our previous work demonstrated that the energy was indeed localized in the space-time power spectrum density $E(\mathbf{k},\omega)$ in the vicinity of the linear dispersion relation~\cite{Cobelli,epjb}. Due to weak interaction between waves, the observed dispersion relation is slightly shifted from the linear one. The observed dispersion relation shows also a finite width 
$\Delta\omega$ that can be attributed to two distinct causes: the nonlinear widening $\Delta \omega_{NL}=1/T_{NL}$ as expected from the WTT and the finite resolution due to the finite size of the measurement picture  $\Delta \omega_{\mathrm{win}}= 2ck\Delta k=2ck\pi/l$. These phenomenon are independent. Hence the effective width is expected to be close to:
\begin{equation}
(\Delta\omega)^2\eqsim(\Delta \omega_{NL})^2+(\Delta \omega_{\mathrm{win}})^2=\frac{A}{T_{NL}^2}+Bk^2
\label{reldisp}
\end{equation} 
where $A$ and $B$ have been substituted to the above theoretical value to account for geometrical constants in the experimental evaluation of $\Delta \omega$ and in the effective shape of the energy profiles due to both phenomena. $A$ is expected to be close to 1 and $B$ of the order of $(2\pi c/l)^2\eqsim65$. 
The integrated over angles power spectrum $E(k,\omega)$ is fitted with a local Gaussian along $\omega$ at given $k$  to measure the width $\Delta\omega(k)$ (as in \cite{epjb}). We compare this width with the nonlinear time measured from the wavelet analysis presented in this paper. According to equation~\eqref{reldisp}, a linear relation is expected between $(\Delta\omega/k)^2$ and $(T_{NL}k)^{-2}$. This is indeed observed in fig.~\ref{large}(a). The extracted values $A$ and $B$ are used to compare $\Delta \omega$ and $T_{NL}$ as a function of $k$ in fig.~\ref{large}(b) and the agreement is very good. This observation validates the fact that the nonlinear time is responsible for the widening of the dispersion relation as predicted by the WTT (even if the observed scaling of $T_{NL}$ is not the one predicted by the WTT).

%%%%%%%%%%%
\section{Conclusion}
%%%%%%%%%%%

We proceeded to a direct extraction of the nonlinear time scale at various wavenumbers and forcing intensities by studying the temporal correlations of wavelet coefficients. With this technique, we have checked that the scale separation hypothesis $Td\gg T_{NL} \gg T$ is reasonably valid in our system which is indeed weakly nonlinear. This hypothesis is one the main requirements for the validity of the Weak Turbulence Theory. Nevertheless, the scaling of the power spectrum were previously observed to be in disagreement with the theory. It appears also that the cascade does not really proceed to the expected dissipative regime for which $T_{NL}$ is comparable to the dissipative time scale. This is due to the fact that the second major hypothesis of the theory is not valid: the asymptotically large system. Finite size effects lead to discrete modes. This discreteness can be ignored if the spectral widening due to nonlinear energy transfers overcomes the frequency separation of adjacent modes in the $\mathbf k$ space. We checked that this requirement is not fulfilled at weak forcing whereas it is valid at low wavenumber for strong forcing intensities. One would expect a range of large wavelength for which the WTT should be valid. But this range is most likely altered by another sort of finite size effects related to the forcing and previously analyzed~\cite{Miquel}: in the vicinity of the forcing point, the phases of the waves with large wavelengths need some distance to get randomized. The plate is most likely too small for the measurement region to be far enough from the forcing so that to really observe a pure WTT regime. Nevertheless, many qualitative feature of the Weak Turbulence Theory and of laminated turbulence have been observed in this system due to the possibility to perform advanced space-time analyses of the turbulent field made possible by the time resolved profilometry technique~\cite{Cobelli1}.

\begin{acknowledgments}
This work was funded by the french Agence Nationale de la Recherche under grant TURBONDE BLAN07-3-197846. We thank Pierre Chainais for the precious discussions on the properties of wavelets and the interpretations of our results.
\end{acknowledgments}

\bibliography{biblio_PRE}

 \end{document}